
\input harvmac
\tolerance=10000
\def\ie{{\it i.e.}}
\def\half{{\textstyle {1 \over 2}}}

\def\Tr{{\rm Tr}}
\def\half{{\textstyle {1 \over 2}}}
\def\np{Nucl. Phys.}

\hfill {DAMTP/94-19}\break

\null
\vskip 0.3in
{\centerline{POINT-LIKE STATES FOR TYPE 2b SUPERSTRINGS}}
\vskip 1cm
 \centerline{ Michael B.  Green,}
\centerline{DAMTP, Silver Street, Cambridge CB3 9EW, UK\foot{email:
M.B.Green@amtp.cam.ac.uk}}
\nopagenumbers
\vskip 4cm
\baselineskip =12pt
This paper considers  closed-string
states of type 2b superstring theory in which the whole
string is localized at a single point in superspace.
Correlation functions of these (scalar and
pseudoscalar) states possess an infinite number of
position-space singularities inside and on the light-cone as
well as a space-like singularity outside the light-cone.\foot{This
paper was previously produced as preprint QMW-91-02.}

\vfill\eject
\baselineskip=12pt
\footline={\hss\tenrm\folio\hss}
\pageno=1
\nref\selfdual{M.B. Green,
{\it Space-time duality and Dirichlet string theory},
Phys.  Lett. {\bf 266B} (1991) 325.;
{\it Duality, Strings and Point-Like Structure}, Talk
presented at the 25th Rencontres de Moriond, Les Arcs, March
1991, ed., J. Tr\^an Thanh V\^an, (Editions Frontieres, 1991).}
General off-shell string correlation functions are
parametrization-dependent expressions of   little apparent   direct
interest.
However, the class of Green functions
which describe the correlation functions of strings in
BRST-invariant point-like
position eigenstates have  special
properties (\selfdual\ and references therein). A
correlation function with
$M$ external point-like states is defined by a sum over
world-sheets of arbitrary genus with $M$ boundaries on which
the space-time coordinates satisfy fixed Dirichlet
conditions, \ie, each boundary (labelled $B$) is
fixed at a space-time point $y_B$.\foot{The theories
discussed here have orientable world-sheets so that cross-
caps are
not included.}   These
Green functions can be obtained (at least formally) from the
partition function of the usual open string theory (in which
the
boundary conditions are the usual Neumann ones) by a $R\to
1/R$
spatial duality transformation \selfdual.

The Dirichlet boundary correlation functions in critical
bosonic string theory exhibit a rich
space-time singularity structure\nref\greenmore{M.B. Green,
{\it Reciprocal
space-time and momentum-space singularities
for the dual string} \np\
{\bf B116} (1976) 449.} \greenmore.
For example the two-boundary
correlation function (with boundaries at positions $y_1^\mu$
and
$y_2^\mu$) not only possesses a light-cone singularity (at
$y^2\equiv (y_2- y_1)^2 =0$), as in conventional point-
particle
Green functions, but there is an infinite sequence of
singularities
inside the light-cone ($y^2 < 0$) as well as a single
space-like
singularity ($y^2>0$).  This illustrates a sort of  duality
between position space and momentum space with the positions
of the $y^2$ singularities
corresponding to the positions of the usual singularities in
momentum
space (with $\alpha'$ replaced by $1/\alpha'$).  The presence of
a space-like singularity is a consequence of the exponential
degeneracy of massive closed-string states (and should therefore
be expected in any string theory) and is closely related to the
open-string
tachyon state --- a familiar feature of  bosonic strings.

The following  will describe  BRST-invariant  point-like states of
ten-dimensional
superstring theory  and their correlation functions.  The
discussion will be given first
in the formalism with world-sheet supersymmetry. The space-
time properties of the point-like states will become clearer
in the manifestly space-time  supersymmetric light-cone
formalism.  It will be seen  that the states are not only
point-like eigenstates of the position operator but also
satisfy (at fixed $\tau$) $\theta^a(\sigma,\tau) |\ \rangle
=0$ or $\partial/\partial  \theta^a(\sigma,\tau)  |\
\rangle=0$, which are natural extensions of the point-like
conditions to the superspace Grassmann spinor coordinate
$\theta^a(\sigma,\tau)$.   These conditions define two
distinct scalar point-like states,  corresponding to off-
shell continuations of  complex combinations of  the two
massless scalar states of type 2b supergravity.   For
certain (`supersymmetric') correlation functions the GSO
projection leads to a cancellation of the space-like
position-space singularity that is present in the individual
spin structures that contribute to the correlation
functions.    This cancellation is a position-space analogue
of the cancellation of the momentum-space tachyon in
superstring theories.  However, the propagator (and  the
commutator) of  closed-string fields in point-like states
possesses  a space-like singularity that can again be related to
an open-string tachyon (in a spin structure that breaks
supersymmetry).

To begin, recall some properties of the bosonic theory in
the
presence of world-sheet boundaries.   In the usual theory
the space-time coordinates $X^\mu$ satisfy Neumann boundary
conditions
$\partial_n X^\mu|_B =0$ (where the subscript $n$ denotes
the
derivative normal to the boundary, $B$).  In this case the
theory
describes interacting open and closed strings and  the
boundary
represents the trajectory of an end-point of an open string.
In the theory with Dirichlet boundary conditions,
$\partial_t X^\mu|_B =0$ (where $t$
denotes the derivatve tangential to the boundary)  each
boundary represents a closed string localized at a  single
(and arbitrary) space-time point, $y_B$.  The partition
function in this  \lq Dirichlet' theory describes the
position-space correlation functions between these point-
like boundary states \selfdual.

The two kinds of boundary condition are formally related by
a target-space duality transformation.   Recall that closed
string theory   compactified on a flat torus with radii
$R^\mu$ ($\mu = 0, \dots 25$)  is equivalent to the
Dirichlet theory on the dual torus with radii $R^{'\mu}
\equiv \alpha' /R^\mu$ and with the integer Kaluza--Klein
charges (labelling the discrete momenta) interchanged with
the winding numbers.  Equivalently,  this transformation
replaces  $X^\mu$ by the dual coordinates, $Y^\mu$, defined
by $\partial_\alpha Y^\mu = \epsilon_{\alpha\beta}
\partial^\beta X^\mu$.   In the presence of world-sheet
boundaries the  Neumann boundary condition on $X$ is
therefore equivalent to  the Dirichlet condition on $Y$
\nref\polchins\polchins{J. Dai, R.G. Leigh and J.
Polchinski, {\it New connections between string theories},
Mod. Phys.  Lett. A vol. 4, {\bf 21} (1989) 2073.}\polchins.
The
Neumann Minkowski-space theory is recovered in the limit
$R\to \infty$ while the Dirichlet Minkowski-space theory is
obtained in the limit $R'\to \infty$ ($R\to 0$).

The usual theory involves a sum over all possible insertions of
boundaries with Neumann boundary conditions in world-sheets of
arbitrary topology while the theory obtained from this by
space-time
duality is one with a sum of all possible insertions  of
Dirichlet
boundaries in world-sheets of arbitrary topology.  If the
sum is taken to include
 integration over the space-time location of each
boundary (so that no momentum flows through it)
\nref\greenbose{M.B. Green, {\it Point-like structure and
off-shell dual strings}, Nucl.  Phys. {\bf B124} (1977) 461;
{\it Modifying the
bosonic string vacuum}, {Phys. Lett.} {\bf 201} (1988)
42.}\greenbose\ the result is a string theory with radically
different properties from those usually considered.

A world-sheet with one or two boundaries may be parametrized
as a cylinder with its axis in the $\tau$ direction (where
the angular coordinate is $\sigma$) and a boundary is
defined
by an end-state,
$|B\rangle$ \nref\callanb{C.G.
Callan, C.
Lovelace, C.R. Nappi and S.A. Yost, {\it Adding holes and
cross-caps to the superstring},
Nucl. Phys. {\bf B293} (1987) 83.}\nref\polchina{J.
Polchinski and Y. Cai, {\it Consistency
of open superstring theories},  { Nucl. Phys.} {\bf
B296}
(1988) 91.} (as in \callanb, \polchina).
When the cylinder ends in a Neumann boundary (on which
$\partial
X^\mu/\partial
\tau = 0$) the end-state  satisfies
$(\alpha_n^\mu +\tilde \alpha_{-n}^\mu)|B\rangle = 0$ (where
the  left-moving and right-moving normal modes satisfy
$[\alpha^\mu_m,\alpha^\nu_n]= m \delta_{m+n}\eta^{\mu\nu} =
[\tilde\alpha^\mu_m,\tilde \alpha^\nu_n]$)
so that the total momentum, $k^\mu \equiv
\sqrt{2/\alpha'}\alpha^\mu_0=
\sqrt{2/\alpha'}\tilde \alpha^\mu_0$, vanishes due to the
$n=0$
condition.
The conditions on a Dirichlet boundary state localized at
$X^\mu = y^\mu$,
are
$$(\alpha_n^\mu - \tilde \alpha_{-n}^\mu)| B,y\rangle=0,
\eqnn\dircon\eqno\dircon$$
for all $n$ (where cylinder end-states of the Dirichlet
theory are denoted by a hat).  A corresponding point-like
eigenstate of the total momentum is defined by
$| B,k\rangle
\equiv \int d^Dy e^{ik \cdot y}| B,y\rangle/(2\pi)^{D/2}$.
A BRST formulation involves ghost coordinates,
$b(\sigma,\tau)$ and
$c(\sigma,\tau)$ and the ghost modes satisfy the boundary
conditions
$$(c_{n} +\tilde c_{-n})| B,k\rangle = 0 = (b_{n} - \tilde
b_{-n})
| B,k\rangle, \eqnn\ghostboun\eqno\ghostboun$$
(where $\{b_n,c_m\}=\delta_{m+n} = \{\tilde b_m,\tilde c_m
\}$)
as they do in the Neumann theory.
These conditions ensure BRST invariance of the state,
$$Q_{BRST}| B, k\rangle=0.\eqnn\brst\eqno\brst$$
\vskip 0.3cm

\noindent{\it Open superstring theory with Dirichlet
conditions.}\par\vskip
0.1cm\nobreak
In type 2b superstring theory (world-sheet) Majorana--Weyl
fermionic spinors,
$\psi^\mu$ and $\tilde \psi^\mu$ (of opposite chirality),
are
introduced with  various combinations of boundary conditions
(spin
structures).   The sectors containing closed-string scalar
bosons
which couple to the end-states of the cylinder are the
anti-periodic $A$ sector (the
$NS \otimes NS$ sector) and the periodic $P$ sector (the
$R\otimes
R$ sector).  When the end-state of a cylinder is either a
Neumann
or Dirichlet
boundary these fields satisfy the boundary
conditions (using conventions which approximate to those of
\polchina)
$$\psi^{\eta\mu}(\sigma,l) \equiv
(\psi^\mu(\sigma,l) + i \eta\tilde \psi^\mu
(\sigma,l))/\sqrt{2}=0,
\eqnn\psieqq\eqno\psieqq$$
where $\eta =\pm 1$.  These boundary conditions on the
spinors in
the cylinder channel are related to the boundary conditions
at the
ends of an open string, $\psi =\pm \tilde \psi$, by a global
diffeomorphism that
rotates the coordinate system through $\pi/2$ --- leading to
the
factor of $i$
in \psieqq\ \polchina.   These translate into the conditions
on the two possible
point-like  endstates,
$$\psi^{\eta\mu}_k| B,\eta,k\rangle \equiv
(\psi_k^\mu + i\eta \tilde \psi^\mu_{-k})| B,\eta,k\rangle
/\sqrt{2}
 \eqnn\modestat\eqno\modestat$$
(where the fermionic modes satisfy the anticommutation
relations
$\{\psi_k^\mu,\psi_l^\nu\} = \eta^{\mu\nu} \delta_{k+l}$).
The index $k$ takes half-integral values in the $A$ sector
and
integral values in the $P$ sector.  These boundary
conditions will
ensure local supersymmetry and BRST invariance.  The
same considerations also determine the boundary conditions
for the
superconformal ghost coordinates,
$\beta,\tilde \beta,\gamma,\tilde\gamma$ (satisfying
$[\gamma_k,\beta_l] =
\delta_{k+l} = [\tilde\gamma_k,\tilde\beta_l]$).

Including all the modes, the Dirichlet boundary states have
the
form
(with an arbitrary overall constant normalization)
$$\eqalign{| B,\eta,k\rangle_{A,P}
= \exp\left(\sum_{n=1}^\infty \right.&\left.
(\alpha_{-n}\cdot \tilde \alpha_{-n}/n - b_{-n} \tilde
c_{-n} - \tilde b_{-n} c_{-n} )\right. \cr
& \left. - i\eta\sum_{k\ne 0} (\psi_{-k} \cdot
\tilde\psi_{-k}
+ \beta_{-k} \tilde \gamma_{-k} + \tilde\beta_{-k}
\gamma_{-k})\right) |0,\eta,k\rangle_{A,P},
\cr}\eqnn\fermstate\eqno\fermstate$$
where $\langle
0,\eta',k'|0,\eta,k\rangle=\delta_{\eta+\eta',0}\delta^{10}(
k+k')$
with $\langle 0,\eta,k| \equiv |0,-\eta,k\rangle^\dagger$
(and the dependence on ghost zero modes has been
suppressed).
The analogous states in the Neumann theory,
$|B,\eta\rangle$, differ only by the sign of $\tilde
\alpha$,
$\tilde \beta$ and $\tilde \gamma$ in the exponent
(and have zero momentum). In the $A$ sector the ground
state,
$|0,\eta,k\rangle_A$, is the tensor
product of two negative $G$-parity $NS$ scalar tachyon
states.  In
the $P$ sector the two ground states, $|0,\eta,k\rangle_P$,
are the states
annihilated by $\psi^{\eta\mu}_0 \equiv \partial/ \partial
\psi_0^{-\eta\mu}$ ($\eta=\pm 1$).    In either sector the
states satisfy the fermionic world-sheet supersymmetry
gauge conditions
$$F_k^\eta| B,\eta,k\rangle_{A,P} \equiv (F_k +i\eta \tilde
F_{-k})| B,
     \eta,k\rangle_{A,P}/\sqrt{2} =0,
\eqnn\fgauge\eqno\fgauge$$
(where $F_k = \sum_n \alpha_n \cdot \psi_{k-n}+\ ghost\
terms$  and $\tilde F_k
= \sum_n \tilde \alpha_n \cdot \tilde \psi_{k-n}+\ ghost\
terms$)
which can be iterated to give the Virasoro conditions,
$(L_n-\tilde L_{-n})| B,\eta,k\rangle_{A,P}=0$.  The
conditions \fgauge\ apply for arbitrary momentum, $k^\mu$
(whereas
the usual
Neumann boundary state satisfies the same conditions only if
$k^\mu=0$).
Using the usual expression for the BRST charge on a
cylindrical
world-sheet
it is easy to see that \fgauge\ leads to $Q_{BRST}|
B,\eta,k\rangle =0$.

In order to simplify the following it is convenient to
specialize to the light-cone gauge, in which the non-
transverse components of the modes of $\psi^\mu$ and $X^\mu$
cancel the ghost modes and only the transverse components of
the modes ($\psi_n^i$, $\tilde \psi_n^i$, $\alpha_n^i$ and
$\tilde\alpha_n^i$ with $i=1, \cdots 8$) survive.

Physical states in the cylinder are obtained by projection
with the usual $GSO$ operator
$P_{GSO} = (1 + \omega(-1)^F)(1+\omega (-1)^{\tilde F})/4$,
where $\omega=1$ in the $A$ sector while $\omega = \pm 1$ in
the $P$ sector (the sign ambiguity corresponding to the two
possible space-time chiralities) and the operators $(-1)^F
\equiv \prod_k (-1)^{F_k}$ and $(-1)^{\tilde F} \equiv
\prod_k (-1)^{\tilde F_k}$ anticommute  with all the
$\psi^i_k$ and $\tilde \psi^i_k$, respectively.  The
Dirichlet
end-states satisfy
$$(-1)^F| B,\eta,k\rangle_A=(-1)^{\tilde F}|
B,\eta,k\rangle_A
 =  -| B, -\eta, k\rangle_A, \eqnn\satisfy\eqno\satisfy$$
in the $A$ sector and
$$ (-1)^F| B,\eta,k\rangle_P  =  (-1)^{\tilde F}|
B,\eta,k\rangle_P=  | B,-\eta,k\rangle_P,
\eqnn\chiral\eqno\chiral$$
in the $P$ sector,
where the property $(-1)^{F_0} |0,+,k\rangle_P
=|0,-,k\rangle_P$
has been used (and $(-1)^{F_0}\equiv \psi^1_0\dots
\psi_0^8$).
[The discussion of these conditions follows closely that
given in
\polchina\  for the usual Neumann boundary.]

The amplitude coupling $M$ on-shell
closed-string states on a world-sheet with a single boundary
(the disk diagram) contains intermediate open strings that
couple to closed strings (which are singlets of the Chan--
Paton symmetry).   Thus, in the usual bosonic  theory with a
$U(n)$ Chan--Paton symmetry the massless gauge-singlet
open-string vector state (which is absent for the
other symmetries allowed by tree-level unitarity, namely,
$SO(n)$
and $Sp(2n)$) mixes with the massless closed-string anti-
symmetric tensor state in a dynamical \lq Higgs'
mechanism\nref\cremmert{E.  Cremmer and J. Scherk, {\it
Spontaneous dynamical breaking
of gauge symmetry in dual models}, Nucl.  Phys.  {\bf B72}
(1974) 117.}\nref\kalb{M. Kalb and P. Ramond, {\it Classical
direct interstring action}, Phys.  Rev. {\bf D9} (1974)
2273.}\nref\shapirod{J.  Shapiro and C.B. Thorn,
{\it BRST-invariant
transitions between closed and open strings},
Phys. Rev. {\bf D36} (1987) 432.} \cremmert-\shapirod.
Theories with Dirichlet boundaries are
quite different since the open-string vector is not a normal
propagating state but is a Lagrange multiplier field which
leads to an  interesting divergence in the bosonic theory
\nref\greenin{M.B. Green, {\it The Influence of World-Sheet
Boundaries on Critical Closed String Theory}, Phys. Lett.
{\bf B302} (1993) 29.}\greenin.    In
the case of the usual type 1 superstring theories the disk
diagram does not  mix  massless open-string and closed-
string states, irrespective of the Chan--Paton group.  This
is a signal that $U(n)$ groups are
inconsistent\nref\schwarzx{J.H. Schwarz, {\it Proc. of the
Johns
Hopkins Workshop on Current Problems in Particle Theory 6},
Florence, 1982.}   due to a mismatch
between the $N=1$ supersymmetry of the open-string states
and the $N=2$
supersymmetry of the closed-string states \schwarzx.   Since
the open-string sector of the supersymmetric Dirichlet
theory does not contain propagating states the status of
this argument needs to be reassessed in that case.

\vskip 0.3cm

\noindent{\it Boundary correlation functions.
}\par\vskip 0.1cm\nobreak
The correlation function of two Dirichlet boundaries,
together with
an arbitrary
number of closed-string vertex operators, defines off-shell
amplitudes of
the type 2b theory coupling two currents and $M$ on-shell
particles.  As a special example consider the two-boundary
correlation function with $M=0$, \ie, the correlation
between two
point-like states at $y_1^\mu$ and $y_2^\mu$, each of which
carries a label $\eta_1$ and $\eta_2$, respectively.  The
complete Green function,
$G_{\eta_1,\eta_2}(y_1-y_2)$,
decomposes into a sum over four spin structures.  When
$\eta_1=\eta_2=\eta$ (the `supersymmetric' case) this sum is
defined by
$$ G_{\eta,\eta}(y_1-y_2) = G^{(++)} + G^{(+-)} + G^{(-
+)}+G^{(--)}, \eqnn\spinstru\eqno\spinstru$$
where the superscripts label the antiperiodic ($+$) and
periodic ($-$) boundary conditions on the fermions in the
$\sigma$ and $\tau$ directions, respectively.  The two-
boundary correlation function in the case $\eta_1=-
\eta_2=\eta$ defines the `propagator' for the point-like
state labelled $\eta$.  The relative signs of the spin
structures in this case differ from \spinstru\ in certain
crucial respects in a manner determined by \satisfy\ and
\chiral.     There are two useful ways of expressing the
Green function, which will now be described.

In the first way the process described by the Green function
is
viewed as the evolution of a closed string from an initial
boundary
state at $y_1$  to the final state at $y_2$, thereby
defining a
cylindrical world-sheet of length $l$.  In this case,
$G_{\eta_1,\eta_2}(y_1-y_2)
= G_{\eta_1,\eta_2}^A(y_1-y_2) + G_{\eta_1,\eta_2}^P(y_1-
y_2)$, where
$$\eqalign{G_{\eta_1,\eta_2}^A(y_1-y_2) &=\int_0^\infty
dl\ {_A\langle}  B, \eta_1,  y_1|(1 + (-1)^F)
e^{-(L_0+\tilde L_0 -1)l} | B, \eta_2, y_2\rangle_A \cr
& = \eta_1\eta_2 \left(G^{(++)} + G^{(+-)} \right)\cr}
\eqnn\acorr\eqno\acorr$$
(note that the GSO
projection  simplifies since a factor of $(-1)^{\tilde F}$
is
equivalent to
$(-1)^F$ and $(-1)^{F+\tilde F}$ is equivalent to 1).  The
evolution operator involves the closed-string hamiltonian,
$L_0+\tilde L_0 -1$, where the Virasoro generators include
all the bosonic
and fermionic coordinates of the $A$ sector.  The first spin
structure in \acorr\ can be expressed in the form
$$\eqalign{G^{(++)} & = \int_0^\infty dl\ _A\langle B,
\eta ,y_1|e^{-(L_0 + \tilde L_0 -1)
l} | B,\eta ,y_2\rangle_A \cr & =
{1\over 2 (\alpha')^5}\int_0^\infty  {dl\over l^5}  e^{-
(y^2/2\alpha'l - l)}
\prod_{n=1}^\infty {(1 + q^{2n-1})^8 \over (1-q^{2n})^8} \cr
& =
{ \pi\over 2 (\pi \alpha')^5}
\int_0^\infty {dl'\over l'}e^{-l'(y^2/4\pi^2\alpha'-1/2)}
\prod_{n=1}^\infty {(1+w^{n-1/2})^8 \over (1-w^{n})^8}
,\cr}\eqnn\xcorr\eqno\xcorr$$
where $y^\mu=y_1^\mu-y_2^\mu$, $w=e^{-l'}$ and $q=e^{-l}$.
The last line is obtained by a standard modular
transformation $l'=2\pi^2/l$ from the one before. Formally
(\ie, ignoring the $l'=0$ end-point divergence),
this expression has an infinite number of logarithmic
singularities in
$(y_1-y_2)^2$ of the form $\ln ((y_1-y_2)^2/4\pi^2\alpha'
+m/2-1/2)$, where
$m\ge 0$ (which is similar to the bosonic Dirichlet theory).
The existence of a singularity outside the light-cone is
correlated
with the
fact that this spin structure is not consistent by itself.
This singularity will soon be shown to cancel in the case
$\eta_1=\eta_2$  when the contribution of $G^{(-+)}$ (the
parity-conserving spin structure in
$G^P$) is added. Furthermore, the $l'=0$ ($l=\infty$)
divergence will cancel
with a similar term arising from $G^{(+-)}$  (the second
spin
structure in $G_{\eta_1,\eta_2}^A$,  which is given by
\xcorr\ with an extra factor
of $(-1)^F$) for either $\eta_1=\pm \eta_2$.

The expression for the $P$-sector correlation function,
$G^P$, is similar to \acorr\  with the crucial difference
that the correlation function with $\eta_1 = -\eta_2$ has
the same sign as the one with $\eta_1 = \eta_2$ (which
follows by use of \satisfy\ and \chiral) so that
$G_{\eta_1,\eta_2}^P = G^{-+} + G^{--}$.  Furthermore,
integration over fermion zero modes causes $G^{--}$ to
vanish (as in the type 1 theory).

The second way of expressing the Green function is to
represent the
cylindrical world-sheet as an annulus, or a  loop formed by
an open
string with end-points fixed at $y_1$ on one boundary and
$y_2$ on
the other.  This may be evaluated as a trace over open-
string
states circulating around the loop.  As is usual with open-
string
loop amplitudes, the resulting expression should be  related
to the
earlier one by a modular transformation in which the
evolution time
$l$ along the cylinder (in \xcorr) transforms into
where $l'=2\pi^2/l$, the evolution parameter for the open
strings
circulating around the annulus.  A string with fixed end-
points has
the mode expansion,
$$Y (\sigma,\tau) = y_1 + (y_2  - y_1 ){\sigma \over \pi}
     + \sqrt{2\alpha'} \sum_{n\ne 0} {1\over n}
   \alpha_n \sin n\sigma e^{in\tau}
\eqnn\openmode\eqno\openmode$$
(which is the $R\to 0$ limit of the more general expression
in
\selfdual) and the   bosonic part of the open-string
evolution
operator for such an open string is given by
$$L^{y,\alpha}_0 = {(y_1-y_2)^2 \over 4\pi^2 \alpha'}
+ \sum_{n=1}^\infty \alpha_{-n}\cdot \alpha_n
\eqnn\posilo\eqno\posilo$$
(where $y = y_2 - y_1 $).
This is to be added to the usual contributions from the
fermionic
modes and
the ghosts, $L_0^\psi$ and $L_0^{ghosts}$ to give the total
$L_0^y$
The propagator, $1/(L_0^y-a)$,
describing the evolution of the world-sheet of an open
string with
end-points fixed at $y_1$  and $y_2$ has poles in $(y_1-
y_2)^2$
(with $a=0$ in the
$R$ sector and $a=1/2$ in the $NS$ sector). These space-time
poles
in the open-string propagator are the origin of the
singularities
seen in the last line of  \xcorr.  More generally,  the
world-sheet for a correlation function of an arbitrary
number of  Dirichlet boundaries can be constructed by sewing
together three-string vertices with  propagators
representing the evolution of open strings with fixed end-
points.  The space-time singularity structure of these
amplitudes is then simply obtained from the singularities of
these propagators.

{}From this viewpoint the expression \xcorr\ (the first
contribution
to $G_{\eta,\eta}^A$ in \acorr) is obtained from the spin
structure in which
the fermionic field is anti-periodic (NS)
between the open-string endpoints, as well as being anti-
periodic
around the loop,
$$ \hat G^{(++)} = {\pi\over 2 (\pi \alpha')^5}
\int_0^\infty {dl'\over l'} \Tr \left(e^{-(L^y_0-1/2)
l'}\right),
\eqnn\traceopen\eqno\traceopen$$
where the hat indicates that the spin structure  is labelled
in the
frame of the annulus.
The $P$-sector parity-conserving term $G^{(-+)}$, which
transforms into $\hat G^{(+-)}$ under the modular
transformation,
arises from a trace over the other
spin structure in the NS sector of the open string,
$$\hat G^{(+-)}={\pi\over 2 (\pi
\alpha')^5}\int_0^\infty
{dl'\over l'}\Tr\left((-1)^{F_{open}} e^{-(L^y_0-
1/2)l'}\right)
\eqnn\tracegso\eqno\tracegso$$
(which includes a factor $-1$ for terms in the trace with an
odd
number of fermionic open-string oscillators).  When
$\eta_1=\eta_2$  the GSO projection gives the sum of
\traceopen\ and  \tracegso, which projects onto the even-
moded Fock space states of the $NS$ sector of the fixed-
endpoint open-string circulating in the loop. The
singularity outside the light-cone  (at $(y_1-y_2)^2=
2\pi^2\alpha'$) cancels in this sum and the leading
singularity is on the light-cone ($(y_1-y_2)^2= 0$).  This
closely parallels the cancellation of
the tachyonic open-string pole in the usual GSO projection.
When $\eta_2=-\eta_1$ the relative sign of $G^{++}$ and
$G^{-+}$ (\ie, of $\hat G^{++}$ and $\hat G^{+-}$) in the
sum changes and the GSO projection no longer eliminates the
space-like singularity -- open-string states of half-integer
mode number are not eliminated.

Similarly, the second term in \acorr, $G^{(+-)}$, is
proportional
to the trace over the open-string states satisfying periodic
(R)
boundary conditions, $\hat G^{(-+)} \sim \int dl' \Tr(\exp
(-L_0^y
l')/l'$.

The Fourier transform of the above expressions with respect
to $y^\mu$ are simple to evaluate, resulting in the
momentum-space
Green function.
The sum of the two spin structures that contribute to
$\tilde
G^A(k)$ (where the tilde indicates the Fourier transform) is
given
by
$$\eqalign{  \eta_1\eta_2 \tilde G_{\eta_1,\eta_2}^A &=
\tilde G^{(++)}+\tilde G^{(+-)} \cr &
=  \half \int_0^\infty dl  e^{-(\alpha' k^2 -2)
l/2}\left(\prod_{n=1}^\infty{(1+q^{2n-1})^8 \over (1-
q^{2n})^8}
- \prod_{n=1}^\infty{(1-q^{2n-1})^8 \over (1-
q^{2n})^8}\right).\cr}
\eqnn\antigreen\eqno\antigreen$$
The one non-vanishing spin structure in the $P$ sector has
the
momentum-space form,
$$\tilde G^{(-+)}(k)= - 8\int_0^\infty dl e^{-\alpha' k^2 l
/2}\prod_{n=1}^\infty{(1+q^{2n})^8 \over (1-q^{2n})^8}.
\eqnn\periogreen\eqno\periogreen$$
For $\eta_1=\eta_2$ these expressions may formally be
interpreted as the
$R\to 0$ limit of the terms in the cosmological constant of
the
Neumann theory in a toroidal space-time. In that case the
fact that the sum of
\periogreen\ and the first term in \antigreen\   is not
exponentially divergent as $q\to 1$ is another expression of
the
fact that the corresponding sum of position-space terms has
no
singularity outside the light-cone. The complete Green
function (the
sum of \antigreen\ and \periogreen) then vanishes for all
$k$ (equivalently, its Fourier transform vanishes for all
$y$) as a
result of the Jacobi {\it aequatio identica satis abstrusa},
which arises
here as  a statement of the equality of the number of scalar
states
coupling to the boundary in the $A$ sector and in the $P$
sector
at every mass level. For $\eta_1=-\eta_2$ the $P$ sector
reinforces the $A$ sector and there is no cancellation.  In
that case the correlation function possesses contains the
space-like singularity at $y^2=2\pi^2\alpha'$ that is
manifest in $G^{(++)}$ (equivalently, in \traceopen).
\vskip 0.3cm

\noindent{\it Space-time supersymmetry}\par\vskip
0.1cm\nobreak
The point-like states can also be described in a manifestly
supersymmetric formalism in the light-cone gauge.  The two
scalar states of the previous sections are given by,
$$| B, \eta, k\rangle_{lc} =
\exp\sum_{n=1}^\infty\left(\alpha_{-n}^{i}
\tilde \alpha_{-n}^{i}/n  - i\eta S_{-n}^{a} \tilde
S_{-n}^{a}\right) |0,\eta
,k\rangle_{lc},\eqnn\lightc\eqno\lightc$$
where the ground states are tensor products of light-cone
ground
states in each Fock space,
$$|0,\eta,k\rangle_{lc} = (|k\rangle  |i\rangle  |\tilde
i\rangle
-  i\eta|k\rangle  |\dot a\rangle  |\tilde{\dot a}\rangle)/4
,\eqnn\groundlc\eqno\groundlc$$
$\eta=\pm$, and $S_n^a, \tilde
S_n^a$ are the modes of the $SO(8)$ transverse space-time
spinors
(the indices $a$ and $\dot a$ label the two inequivalent
$SO(8)$
spinors
while $i$ labels the vector).  These ground states are the
two
complex scalar
ground states of the type 2b theory.  The states \lightc\
satisfy
$$S^{\eta a}_n| B,k,\eta\rangle_{lc} \equiv
{1\over \sqrt 2}(S_n^a + i\eta \tilde S^a_{-n})|
B,k,\eta\rangle_{lc} =
0 \eqnn\scon\eqno\scon$$
(for all $n$) in addition to the point-like condition on the
bosonic
coordinates.  The ground states \groundlc\ are again
normalized so
that
$_{lc}\langle 0,\eta',k'|0,\eta,k\rangle_{lc} =
\delta_{\eta+\eta',0}
\delta^{10}(k+k')$ (using
$\langle i|j\rangle =\delta_{i,j}$, $\langle a|b\rangle
=\delta_{a,b}$), where $_{lc}\langle 0,\eta,k| = (|0,-
\eta,k\rangle_{lc})^\dagger$.
They are related by
$|0,\eta, k\rangle _{lc} = S_0^{\eta 1}S_0^{\eta 2}\dots
S_0^{\eta 8}|0,-\eta,k\rangle_{lc}$.
The end-state \lightc\ is not an eigenstate of the twist
operator,
$\Omega$, that interchanges the left and right-moving Fock
spaces.  Expanding the exponential in a power series leads
to an
infinite
series of Fock-space states that includes states of the form
$(S^\dagger\tilde S^\dagger)^{2n+1}|i\rangle  |\tilde
i\rangle$
and $(S^\dagger \tilde S^\dagger)^{2n}|\dot a\rangle |\tilde
{\dot
a}\rangle$,
which are antisymmetric under $\Omega$, in addition to the
symmetric states.

The states \lightc\ satisfy the 16-component supersymmetry
conditions
$$Q^{\eta a}| B, \eta,k\rangle_{lc} =0=
Q^{\eta\dot a} |
B, \eta,k\rangle_{lc},\eqnn\supsymm\eqno\supsymm$$
where
$Q^\eta = \left(Q +i\eta\tilde Q \right)/\sqrt{2}$
and
$$Q^a=\sqrt{2k^+}S_0^a, \qquad Q^{\dot a}=\sqrt{2\over
\alpha' k^+}
\gamma^i_{a\dot a}\sum_{-\infty}^\infty S_{-n}^a \alpha_n^i,
\eqnn\superdef\eqno\superdef$$
are the two SO(8) components of a 16-component space-time
supercharge (with analogous definitions for $\tilde Q^a$ and
$\tilde
Q^{\dot a}$).  These supercharges satisfy the usual
superalgebra which includes the relations
$Q^{+\dot a 2} = Q^{-\dot a 2} =  P^- -  \tilde P^-$
and
$\{Q^{+\dot a},Q^{-\dot b}\} = \delta^{\dot a\dot b} (P^- +
\tilde
P^-)$,
where $P^-\equiv H$ is the light-cone gauge hamiltonian.
The supersymmetry conditions, \supsymm, differ from those of
the type 1 superstring by the presence of important factors
of $i$ in the definition of $Q^\eta$ (arising from the
replacement $k^+ \to -k^+$ in the definition of $\tilde Q$
in passing from the Neumann to the Dirichlet theory).

In a light-cone superspace formulation of the type 2b theory
\nref\brinka{M.B. Green, J.H. Schwarz and L. Brink,  {\it
Superfield Theory of Type II Superstrings},
{Nucl. Phys.} {\bf B219} (1983) 437.}\brinka\  a $SO(8)$
Grassmann spinor coordinate can be identified,
$$\theta^a(\sigma,\tau) = {1\over\sqrt{ 2k^+}}
    \left(S^a(\sigma,\tau)+i\tilde S^a(\sigma,\tau)\right),
\eqnn\thetdef\eqno\thetdef$$
together with a conjugate spinor momentum,
$$\lambda^a(\sigma,\tau) =
\sqrt{k^+/2}\left(S^a(\sigma,\tau)
 - i\tilde S^a(\sigma,\tau)\right)
    \equiv \partial/\partial \theta^a(\sigma,\tau).
\eqnn\lambdef\eqno\lambdef$$
The components of the supercharges $Q^{\pm a}$ can be
identified
with
the zero modes,
$Q^{+ a} =\sqrt{2} k^+\theta^a_0$, $Q^{- a}
=\sqrt{2}\lambda^a_0$,
while the supercharges  $Q^{\pm \dot a}$ have simple
representations
as bilinears in $\theta^a$, $\partial/\partial \theta^a$
and $\partial_\pm Y^\mu$ \brinka.
The two states, \lightc, satisfy
$$\theta^a(\sigma,\tau)| B,k,+\rangle_{lc} =0,\qquad
\lambda^a(\sigma,\tau)| B,k,-\rangle_{lc} =
0.\eqnn\lambthet\eqno\lambthet$$
Either of these conditions is an obvious superspace
extension of
the Dirichlet condition, \dircon.

A string light-cone superfield containing these states can
be
written
as an expansion in component string fields,
$$\Phi[Y^i(\sigma),\theta^a(\sigma),k^+] =
\phi[Y^i(\sigma),k^+]
+ \dots +  \phi_{n_1n_2\dots n_m}
^{a_1 a_2 \dots a_m}[Y^i(\sigma),k^+]\theta_{n_1}^{a_1}
\theta_{n_2}^{a_2}\dots \theta_{n_m}^{a_m} +
\dots,\eqnn\expansion\eqno\expansion$$
which is the string generalization of the light-cone
superfield of
ten-dimensional type 2b supergravity \nref\greentt{M.B.
Green and
J.H. Schwarz, {\it Extended Supergravity in Ten Dimensions},
Phys. Lett. {122B} (1983) 143.}\greentt.
The states \lambthet\ are those at the top and bottom ends
of the
infinite-dimensional supermultiplet (annihilated by $Q^{\pm
a}$).
Upon imposing the point-like condition (so that the
component
string fields are
proportional to $\delta(Y^i(\sigma) - y^i)$) these top and
bottom
states are
also annihilated by $Q^{+\dot a}$ and $Q^{- \dot a}$,
respectively
(these
are the non-linearly realized $SO(8)$ light-cone
supercharges) so
that
these states are the top and bottom states of a covariant
physical
supermultiplet.   They therefore have the appropriate
quantum numbers to couple to the complex massless scalar (or
its complex conjugate) of type 2b supergravity, as well as
an infinite sequence of massive scalar states.

Scattering amplitudes can be calculated simply using the
off-shell
states \lightc\ in the Dirichlet theory.
For example, the cylinder amplitude with two Dirichlet
boundaries
(with momentum $p_1$ and $p_2$) and  $M$ massless closed-
string
ground states with momenta $k_r$ (satisfying momentum
conservation,
$p_1+p_2 +\sum_r k_r =0$) attached at points
$\rho_r=(\sigma_r,
\tau_r)$ to the interior is proportional to (setting
$\alpha'=2$
for simplicity)
$$\eqalign{&A(\{k_r\}, p_1,p_2)
= g^M n^2\sum_{perms}  \left(\prod_{r=1}^M \int_0^{2\pi}
d\sigma_r\right)
{_{lc}\langle}  B, \eta_1,p_1|\int_{\tau_M}^\infty dl
e^{-(p_1^2+N+\tilde N)l}
\cr & \quad \int_{\tau_{M-1}}^\infty d\tau_M V_M
e^{-((p_1+k_M)^2 + N+\tilde N)\tau_M}  \dots \int_0^\infty
d\tau_1
V_1 e^{-(p_2^2 + N+\tilde N) \tau_1}
| B, \eta_2,p_2\rangle_{lc},\cr}\eqnn\loopamp\eqno\loopamp$$
where $V_r(\sigma_r,\tau_r,k_r)$ is the light-cone vertex
for the emission of the $r$th on-shell state with
transverse momentum $k_r^i$
and polarization $\zeta_r^{MN}$ (where $M,N$ label vector or
spinor
indices of
the external states) and the sum is over all permutations of
the
ordering of the vertices in $\tau$.
The momenta in this expression are defined in a special
frame in
which $k_r^+=0$, so it will not describe the most general
kinematic
configuration when there are many external particles.
Subject to
this restriction, the amplitude \loopamp\  is the same as
that
obtained in the covariant approach after summing over spin
structures.

The vanishing of the two-boundary amplitude ($M=0$) in the
case $\eta_1=\eta_2$ is here seen to be due to the
integration over fermionic zero modes associated with
space-time supersymmetry.  The propagator for the $\eta_1$
state is the $M=0$ amplitude with $\eta_1=-\eta_2$, which
does not vanish. Using $(-1)^F| B,\eta, p\rangle = | B, -
\eta, p\rangle$ it is clear that the $\tau$ boundary
condition breaks supersymmetry -- the open-string Grassmann
coordinates circulating around the annulus have  $\half$-
integer modes.  The resulting momentum-space propagator has
the form
$$\tilde G \sim \int_0^\infty dl e^{-\alpha' p^2 l
/2}\prod_{n=1}^\infty{(1+q^{2n})^8 \over (1-q^{2n})^8},
\eqnn\suppergreen\eqno\suppergreen$$
which is similar to \periogreen\ and is the Fourier
transform of a position-space expression with a space-like
singularity.

Amplitudes with two Dirichlet boundaries (labelled $\eta_1$ and
$\eta_2$)
and two or more external
on-shell closed-string ground vertex operators  ($M\ge 2$)
do not necessarily vanish even if
$\eta_1=\eta_2$ since each vertex introduces four
fermionic modes.
{}From the earlier  analysis it is evident that they may be
expressed
in terms of  poles and multi-poles in position space.
In order to exhibit these position-space singularities  it
is again convenient to  represent the amplitude as an open-
string loop diagram in which the circulating open string has
end-points fixed at positions $y_1$ and $y_2=y-y_1$,
$$\eqalign{A(\{k_r\}, y_1, y_2)  &  \equiv \int {dp_1\over
(2\pi)^5} {dp_2\over (2\pi)^5} e^{ip_1\cdot y_1 + ip_2 \cdot
y_2}  A(\{k_r\}, p_1,p_2)\delta(p_1+p_2+\sum_{r=1}^M k_r)
\cr &
\sim \Tr \left({1\over L_0^y} V_1{1\over L_0^y} V_2 \dots
{1\over L_0^y} V_M \right),\cr}
\eqnn\loopdia\eqno\loopdia$$
where $L_0^y  = y^2/4\pi^2\alpha' + N_\alpha + N_S$
($N_\alpha$ and $N_S$ are the level numbers in the Fock
spaces of the bosonic and fermionic light-cone coordinates).
In this expression the $y^2$ singularities (poles and
multipoles) arise manifestly from the zeroes of $L_0^y$.
When  $\eta_1=\eta_2$  the open-string Grassmann coordinates
in $L_0$ have integer modes and the open-string supercharge
is well-defined.  In that case the leading
singularity is on  the light cone and there an infinite
number
of  singularities inside
the light-cone, separated by $4\pi^2\alpha'$ \greenmore.
The behaviour
of these amplitudes in the deep inelastic region (\ie, for
large
space-like momenta in the off-shell legs) is  dominated by
the light-cone
singularities, leading to scaling with anomalous dimensions.
The
first singularity inside the light-cone
gives an exponentially suppressed correction to the light-
cone
behaviour.
In the limit $\alpha'\to \infty$ (or string tension $T\to
0$) the singularities
inside the light-cone move to infinity, leaving just the
light-cone
singularity. If $\eta_2=-\eta_1$ the Grassmann coordinates
have half-integer modes, there is no supercharge in the
open-string sector and the leading singularity is at
$y^2=2\pi^2\alpha'$. These considerations generalize the
comments outlined
in \greenmore,\selfdual\ in the context of the bosonic
theory.

\vskip 0.3cm

This article has been concerned with correlation functions
of a very special class of BRST-invariant states, in which
each external string is localized at a point in superspace.
These states are  complex scalar end-states of a string
superfield. The correlation function of a pair of these
states of opposite type (the `propagator') is characterized
by an infinite sequence of position-space singularities
inside and on the light-cone as well as the space-like
singularity at $y^2=2\pi^2\alpha'$, rather as in the bosonic
theory.  Correspondingly, the free propagator (or, equivalently,
the equal-time commutator) possesses a singularity at space-like
separations (which is presumably a sign of further space-like
structure in the propagator of more general string states).
Although  the physical relevance  of this
observation is unclear it may be  connected to questions of
causality in superstring  theory (for example, in the
context of the arguments in \nref\martineca{E.  Martinec,
{\it The Light Cone in String Theory}, EFI-91-23/HEP-TH
9304037.} \martineca). The (`supersymmetric')  correlation
function of two end-states of the same type (\ie, from the
same end of the space-time supermultiplet) vanishes, but the
correlation functions describing the coupling of two end-
states of the same type with two or more on-shell states are
non-vanishing  and exhibit the  striking  feature that the
space-like position-space singularity  cancels.

\listrefs
\bye